\documentclass[a4paper]{article}
\usepackage{ifxetex}
\ifxetex
    \usepackage{gc2019eng}
\else
    \usepackage{gc2019eng_latex}
\fi
\usepackage{hyperref}
\usepackage{subcaption}
\usepackage{graphicx} 
\title{Barriers towards no-reference metrics application to compressed video quality analysis: on the example of no-reference metric NIQE}

\author{A.~Zvezdakova$^1$, D.~Kulikov$^2$, D.~Kondranin$^3$, D.~Vatolin$^4$}

\email{\{azvezdakova|dkulikov|denis.kondranin|dmitriy\}@graphics.cs.msu.ru}

\organization{$^{1,3,4}$Lomonosov Moscow State University, Moscow, Russia;\par $^2$Lomonosov Moscow State University, Moscow, Russia; Dubna State University, Dubna, Russia}

\abstract{This paper analyses the application of no-reference metric NIQE to the task of video-codec comparison. A number of issues in the metric behavior on videos was detected and described. The metric has outlying scores on black and solid-colored frames. The proposed averaging technique for metric quality scores helped to improve the results in some cases. Also, NIQE has low-quality scores for videos with detailed textures and higher scores for videos of lower bit rates due to the blurring of these textures after compression. Although NIQE showed natural results for many tested videos, it is not universal and currently can’t be used for video-codec comparisons.}


\authorsInfo
{
    Anastasia Zvezdakova, PhD student. Received her M.S. degree in computer science from Moscow State University in 2018. Currently, she is a PhD student at Moscow State University and a member of Video Group in MSU Graphics\&Media Lab. Her research interests involve video codecs analysis and optimization, stereoscopic video subjective quality assessment. Anastasia is one of the contributors to MSU Video Codec Comparison Project \url{http://www.compression.ru/video/codec_comparison/index_en.html} and to the 3D video quality measurement project VQMT3D \url{http://compression.ru/video/vqmt3d/}. E-mail aantsiferova@graphics.cs.msu.ru

    Dmitriy Kulikov, PhD in computer science. Received his M.S. degree in 2005 and his PhD in 2009, both from Moscow State University. Currently, he is head of the MSU Video Codec Comparison Project in Video Group at the CS MSU Graphics\&Media Lab. Also, he is an associate professor in Dubna State University. His research interests include compression methods, video processing algorithms, video quality metrics, machine learning. Dmitriy is one of the main contributors to MSU Video Codec Comparison Project \url{http://www.compression.ru/video/codec_comparison/index_en.html}.  E-mail dkulikov@graphics.cs.msu.ru

    Denis Kondranin, student. Received his B.S. degree in 2018 from Moscow State University. Currently, he is an M.S.student and a member of Video Group, a part of MSU Graphics\&Media Lab. His research interests involve video codecs comparison, encoding options analysis and optimization. Denis is one of the contributors to the 3D video quality measurement project VQMT3D. E-mail denis.kondranin@graphics.cs.msu.ru

    Dmitriy Vatolin, PhD in computer science, head of Video Group (CS MSU Graphics\&Media Lab). Graduated from the Computer Science department of Lomonosov Moscow State University in 1996 and got his PhD in 2000. The theme of his PhD thesis was ``Optimization methods of fractal image compression''. Dmitriy has been teaching a computer graphics course at MSU since 1997. He wrote the book Algorithms of Image Compression in 1999 and coauthored Methods of Data Compression in 2003. Dr Vatolin supervised collaborative research projects with Intel, Real Networks and Samsung. He created one of the largest websites devoted to data compression (http://compression.ru). He teaches courses on methods of 3D and 2D video and image processing and compression. E-mail dmitriy@graphics.cs.msu.ru
}

\begin{document}

\maketitle

\begin{multicols*}{2}

\section{Introduction}

Today video content takes the biggest part of world Internet traffic (more than 70\%). According to the forecasts \cite{cisco_2017}, its rate will grow up to 82\% in 2022. This trend leads to the creation of new encoding standards and improvements in existing encoders. There is a number of video-codec comparisons which are conducted to find the best codecs for different tasks and use cases and to help users and customers to find appropriate encoders for their needs. The target for video encoding is to deliver high visual quality with reduced file size, so the only reliable way to compare encoded videos quality is to perform a subjective evaluation. It requires a proofed methodology and a high number of observers to achieve reasonable results. In general, subjective comparisons are still very expensive to perform, however, there are some services which help researchers to perform qualitative subjective comparison \cite{subjectify}. This obstacle increases the importance of objective metrics for video quality comparison. 

Objective quality metrics can be divided into three general categories: full-reference metrics, no-reference metrics and reduced-reference metrics. Full-reference metrics are easy to interpret and useful in application to video compression quality estimation. Unlike full-reference metrics which require source video to compare with compressed, no-reference metrics are useful when you don't have a source and want to estimate the quality of the compressed video. This case is usual for example for cloud encoding when videos are uploaded compressed by a built-in encoder in smartphones or non-professional cameras. Reduced-reference metrics require just some part of information about source video and can also be used in some of the listed cases.

\section{Related work}

There is a number of no-reference metrics which were created using databases with subjective quality scores. Such quality assessment models were trained to estimate subjective quality, and so their scores depend on training and testing sets. For example, DIIVINE (2011) \cite{diivine2011}, LBIQ (2011) \cite{lbiq2011}, BRISQUE (2012) \cite{brisque2012} and V-Bliinds (2012) \cite{bliinds2012} were trained on LIVE data set. 
In 2015, a metric called IL-NIQE \cite{il_niqe_2015} was proposed. It was based on NIQE \cite{niqe} metric, which is studied in this paper, but used multivariate Gaussian (MVG) model to predict the quality of image patches instead of using a single global MVG model for an image. 

Another group contains metrics which weren't trained on any data sets and use only data from a source image to estimate its quality. For example, CORNIA (2012) \cite{cornia} combined feature and regression training. Recently several approaches which use neural networks architectures have been developed. The authors of COME (2018) \cite{come_2018} proposed an approach based on convolution neural network AlexNet and multi-regression which outperformed V-Bliinds on a number of video sets. 

No-reference metrics are created to approximate users perception of video quality, but in case of estimating the quality of encoding and compression, they can be used only as an addition to reference metrics. No-reference metrics can't become the main criteria for encoders comparison because in the opposite way encoders could win the comparison producing a visually ideal result which has little common with the input video. The authors of this paper organize worldwide video-codec comparisons for 16 years. Currently, full-reference metric SSIM is used in these comparisons as the main metric supplemented with a number of additional metrics (PSNR, VMAF). At the same time, several researchers and industry experts consider measuring and taking into account no-reference metrics in video-codec comparisons. This paper describes the authors' experience of using no-reference metric NIQE (Natural Image Quality Evaluator) \cite{niqe} created by Anish Mittal, Rajiv Soundararajan and Alan C. Bovik in video-codec comparison. This metric is one of the most popular nowadays and shows good results for image quality assessment.

We used NIQE to access the quality of encoded video sequences during the video-codec comparison. The main idea of NIQE metric is based on constructing a collection of quality-aware features and fitting them to a multivariate Gaussian (MVG) mode. NIQE score represents the degree of distortions in the frame, and the lower score is, the higher quality is the frame. Accordingly, rate-distortion graphs for encoded videos look unusually inverted, so on the plots in this paper NIQE scores are presented inverted to make the results more familiar and interpreting.

There is an open implementation on MATLAB provided by the authors \cite{niqe_matlab}. In order to increase computational speed, we used an implementation from MSU Video Quality Measurement Tool (VQMT) which is currently faster. The tool has a free version (it includes NIQE) and can be downloaded \cite{vqmt}. Speed was important in this case because the metric was used for video quality assessment.                                     

\section{Experimental setup}

For the evaluation, 28 different FullHD video sequences were used with number of frames per second from 24 to 60 and which were generated by real users. The videos were chosen from MSU video collection which consists of 15,833 videos. The collection was divided into 28 clusters by spatiotemporal complexity \cite{google_si_ti} and one sequence from each cluster, which was close to the cluster center, was chosen for the final testing set. 
Each video was encoded by x264 and x265 encoders. There were three encoding use cases (``fast'', ``universal'' and ``ripping'') based on different encoding speed/quality ratios and 7 different bit rates from 1 Mbps to 12 Mbps. An overall number of encoded streams which were evaluated by NIQE is 1176.

The final video set was used in 2018 Moscow State University (MSU) video-codec comparison \cite{msu_comparison_2018}. The comparison results are available on the link, but the results of NIQE were not published on-line because of several issues found in NIQE application to video quality measurement. Some of them were noted it the original article, the others were resolved with our proposed averaging technique which will be described in the article. Unfortunately, some issues can't be fixed without the metric improvement (completing the training set or other fixes). In this article, we suggest the method of metric results processing to solve the detected problems on metric application to videos.

\section{Metric behavior on videos}

For most of the encoded videos, NIQE showed the results which reflected the usual perceptual video quality on different bit rates. But there were some cases in which NIQE showed the results with some issues; the following sections describe the detected issues and their reasons.

\subsection{Cases with relevant results}

According to the authors, NIQE is not applicable to unnatural distortions in scenes and scenes from unnatural source (e.g. computer graphics), as such scenes were not used during the training. However, we checked metric scores on cartoons from our video set. 

\begin{figure}[H]
     \begin{subfigure}{.23\textwidth}
     \includegraphics[width=\linewidth]{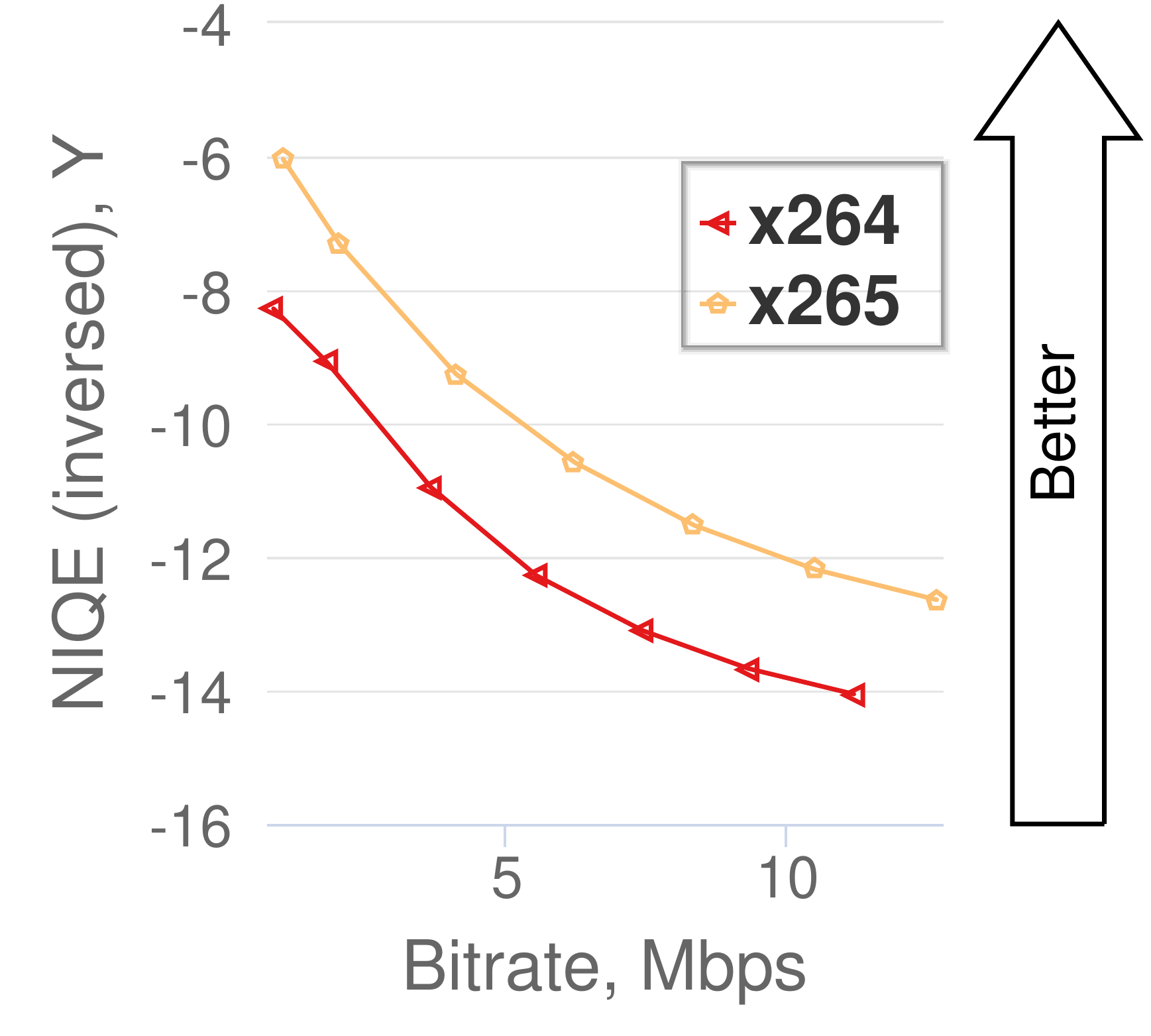}
     \caption{\textit{Sita} video sequence} 
     \label{fig:rd_sita}
     \end{subfigure}
     \hspace*{\fill}%
     \begin{subfigure}{0.23\textwidth}
     \includegraphics[width=\linewidth]{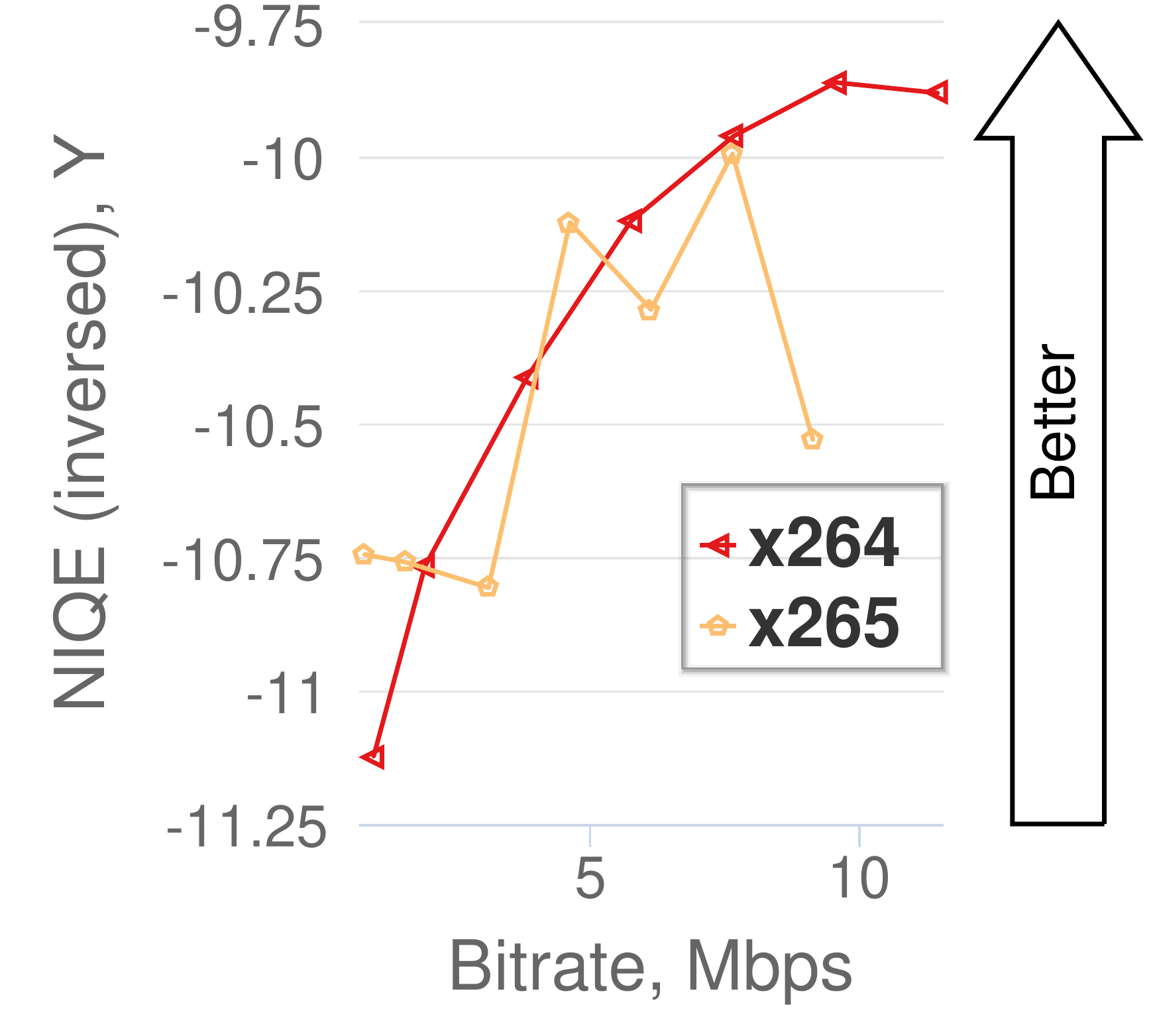}
     \caption{\textit{Sintel} video sequence} 
     \label{fig:rd_sintel}
     \end{subfigure}
     \caption{Rate-distortion graph for animation.} 
     \label{fig:synthetic}
 \end{figure}

At \textit{Sita} (part from the cartoon movie), rate-distortion curve looks inverted (Fig.~\ref{fig:rd_sita}), NIQE shows worse quality scores for high bit rates that for low bit rates. This means that the metric is really not applicable to this type of content.
At \textit{Sintel} (part from CGI movie trailer), NIQE showed non-monotonic scores for x265 encoder on fast use case bit rate map, but acceptable results for universal and ripping use cases (Fig.~\ref{fig:rd_sintel}).
Thus, the metric is said to be not applicable to cartoons, but we revealed that it works for some types of realistic animation, such as for video gaming (sequences \textit{Witcher3}, \textit{Rust}).

\begin{figure}[H]
    \includegraphics[width=.98\linewidth]{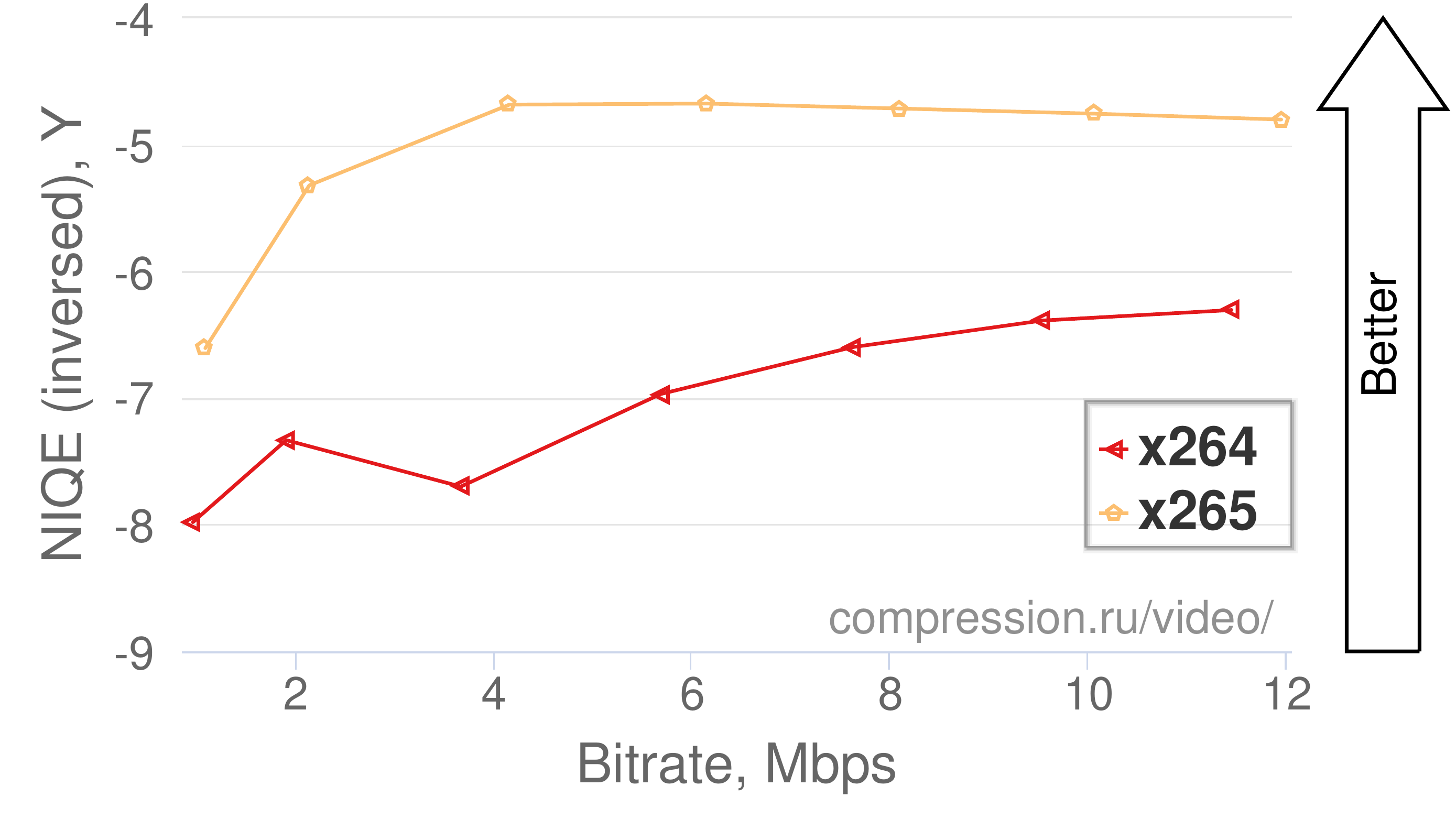}
    \caption{Rate-distortion graph for \textit{Hera}.}
    \label{fig:hera}
\end{figure}

\begin{figure}[H]
     \begin{subfigure}{.98\linewidth}
     \includegraphics[width=\linewidth]{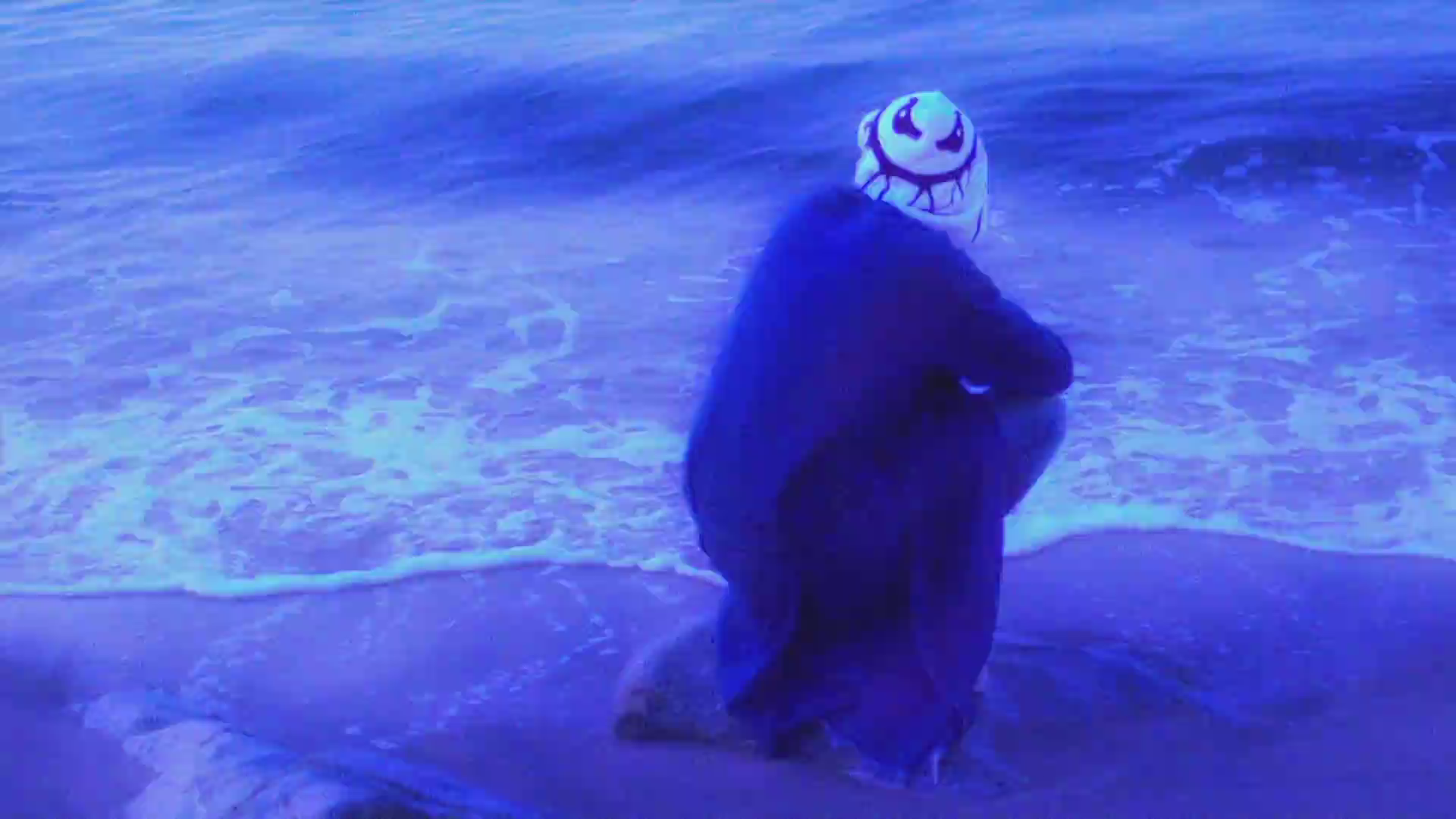}
     \end{subfigure}
     \\
     \begin{subfigure}{.98\linewidth}
         \begin{subfigure}{.49\linewidth}
         \includegraphics[width=\linewidth]{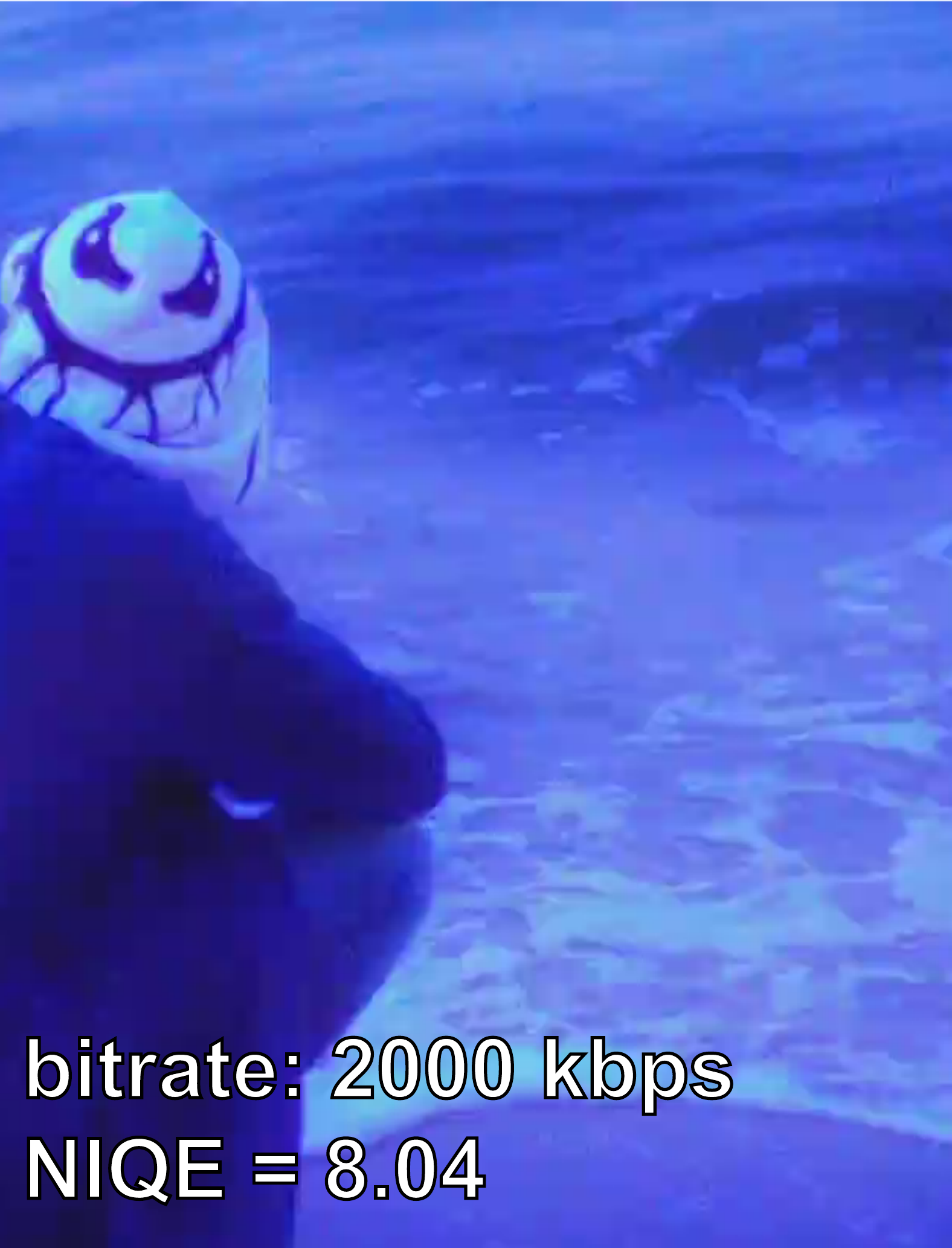}
         \end{subfigure}
         \hfill
         \begin{subfigure}{.49\linewidth}
         \includegraphics[width=\linewidth]{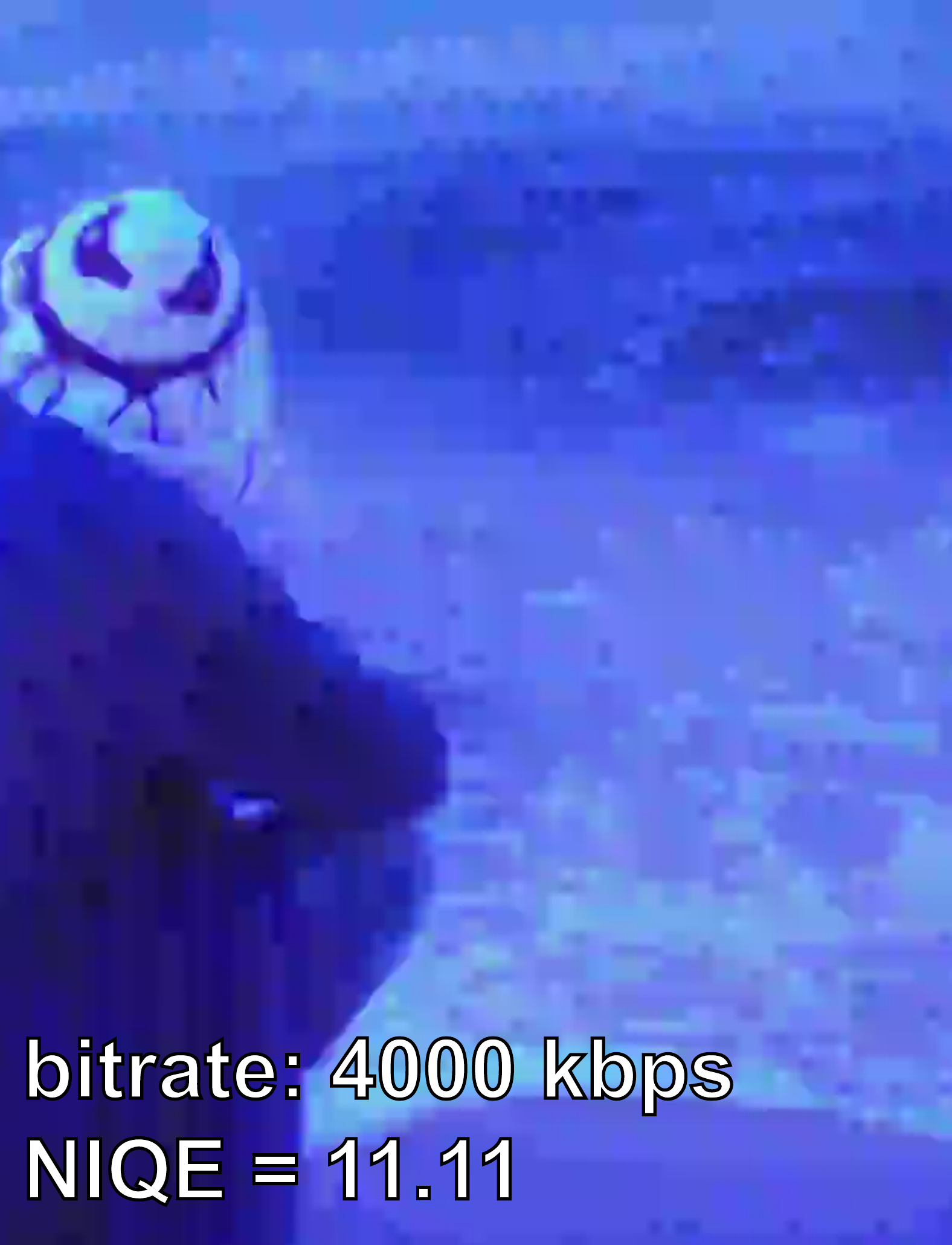}
         \end{subfigure}
     \end{subfigure}
     \caption{Frame 208 from \textit{Hera} video sequence, codec: x264, fast use case. According to NIQE, left image is visually better.}
     \label{fig:hera_vis}
 \end{figure}

There were some examples, where the rate-distortion curve looked unnatural, but the metric correctly ranked worse visual quality to higher bit rates. For example, on \textit{Hera} video sequence (a part of a music clip with grain effects) NIQE showed worse score for x264 encoding on 4000 kbps than on 2000 kbps in fast use case (Fig.~\ref{fig:hera}). The metric had better scores for almost all frames of the lower bit rate. It is shown in the example frame on Fig.~\ref{fig:hera_vis}, where x264 encoding of the video on 4000 kbps produced worse visual quality and more compression artifacts than on 2000 kbps.

\subsection{Cases with irrelevant results}
\subsubsection{Dark scenes}
The metric was said to be not applicable to the cartoons, but some other types of video content also had inaccurate NIQE scores. One of the most frequent cases in video sequences with completely black frames (for example, in the beginning). These frames, according to NIQE, are perceptually worse than the other frames and has an extremely high metric score. This might happen because of the absence of such kind of content in training data used for NIQE creation.

\begin{figure}[H]
    \begin{minipage}[b]{.98\linewidth}
    \centering{\includegraphics[width=\linewidth]{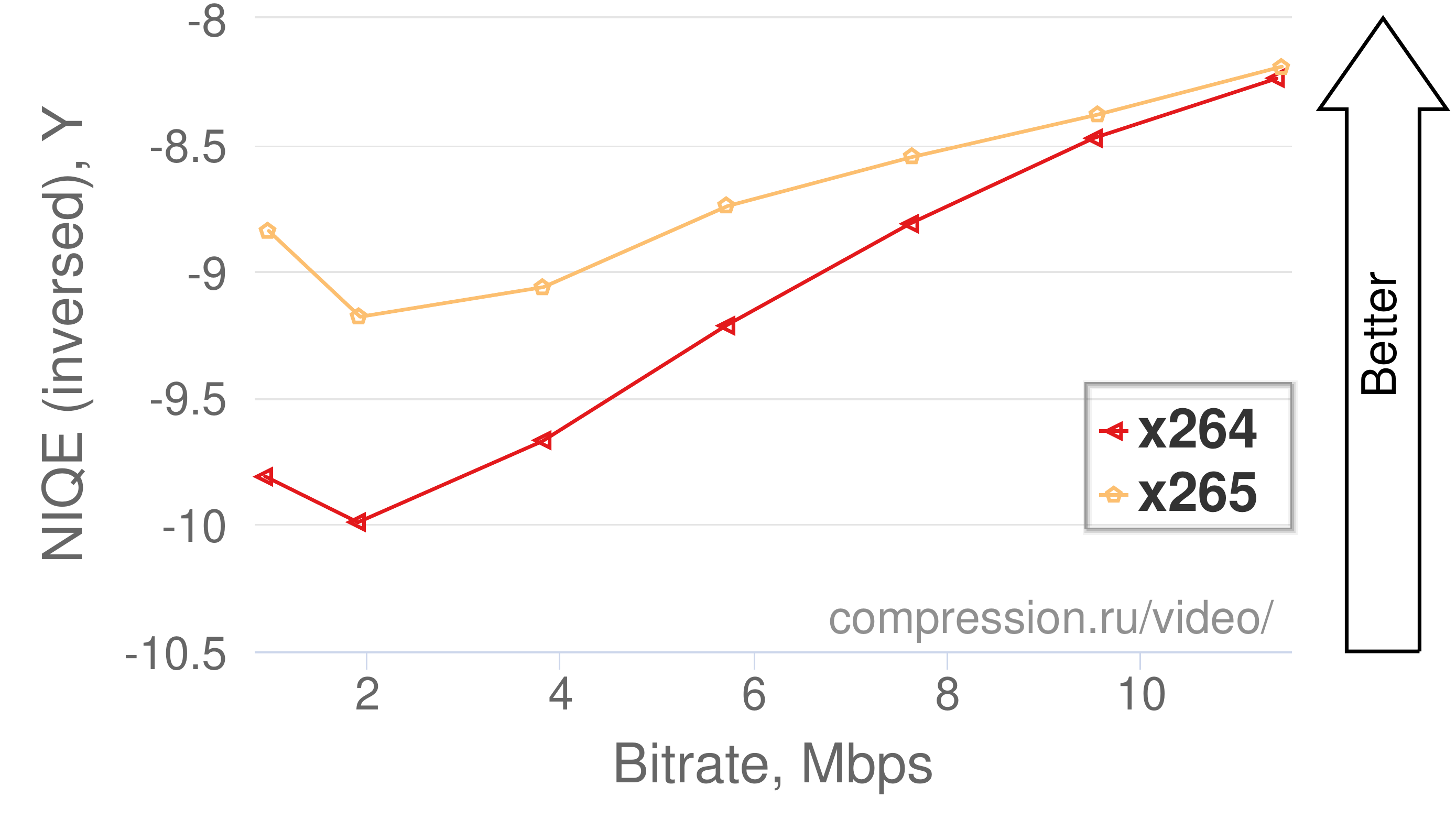}}
    \caption{Rate-distortion graph for \textit{Fire}.}
    \label{fig:fire}
    \end{minipage}
\end{figure}

For example, for x264 encoding NIQE showed worse score on 2000 kbps than on 1000 kbps at \textit{Fire} video sequence (Fig.~\ref{fig:fire}). It contains close shooting of a fire in a dark. In this sequence, the metric showed better scores on a group of frames where the camera started a slow movement.

\begin{figure}[H]
    \begin{subfigure}{.23\textwidth}
    \includegraphics[width=\linewidth]{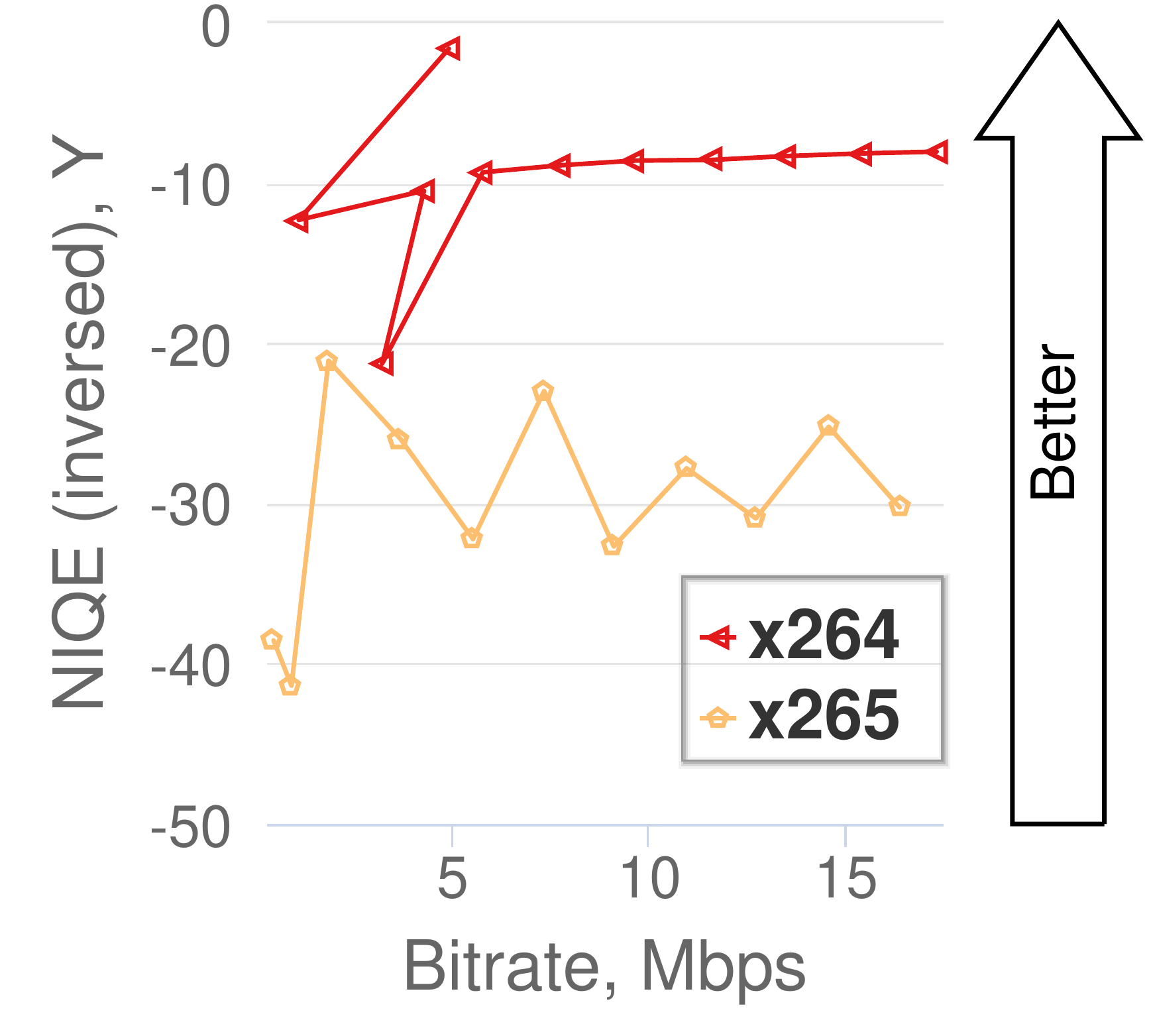}
    \caption{Rate-distortion graph} \label{fig:music_clip_rd}
    \end{subfigure}
    \hspace*{\fill} % separation between the subfigures
    \begin{subfigure}{0.23\textwidth}
    \includegraphics[width=\linewidth]{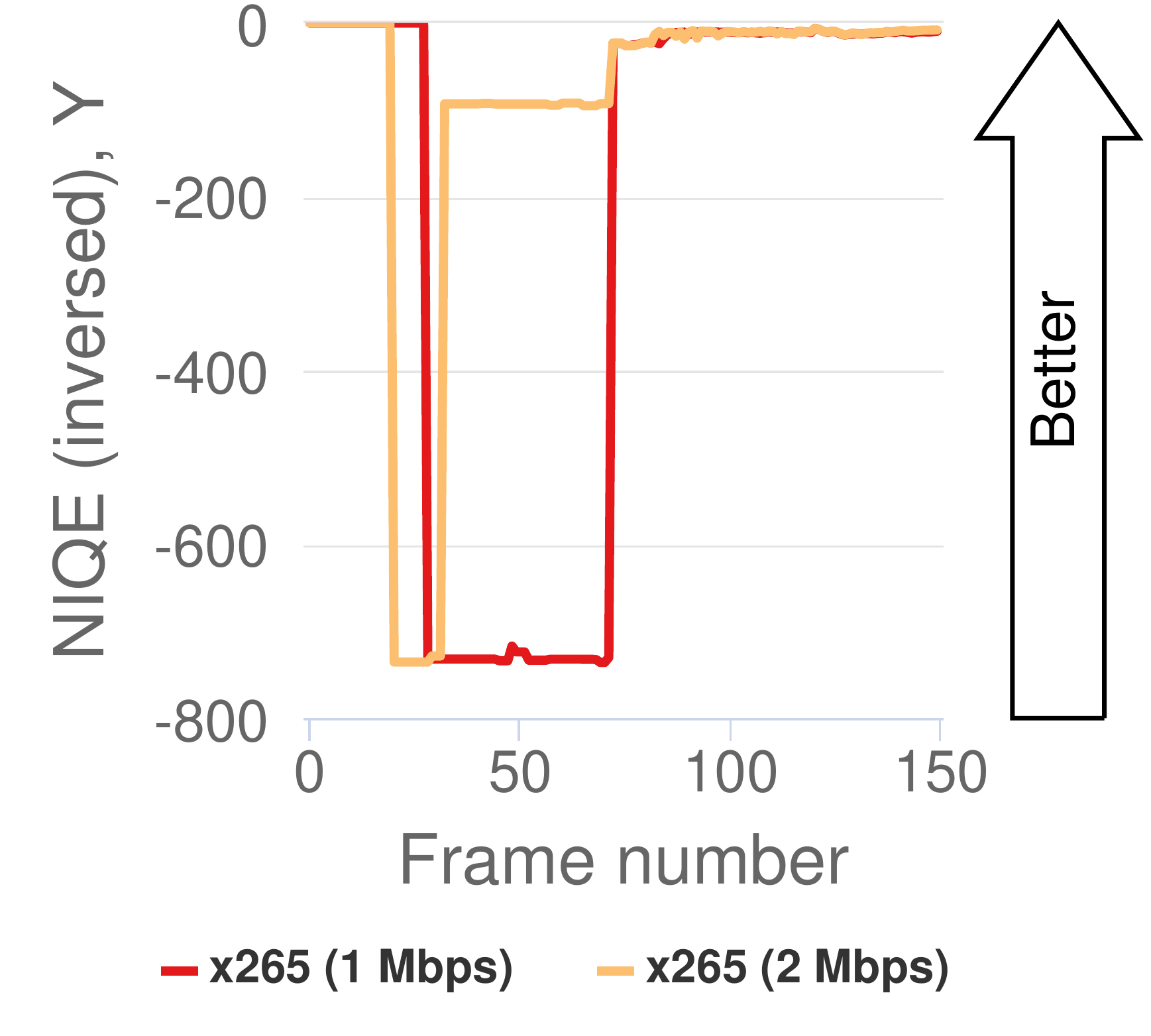}
    \caption{Per-frame NIQE scores} \label{fig:music_clip_per_frames}
    \end{subfigure}
    \caption{\textit{Music clip} video sequence.} 
    \label{fig:music_clip}
\end{figure}

Another example which demonstrates this issue is presented in Fig.~\ref{fig:music_clip}. \textit{Music Clip} video sequence was quite complicated for many encoders in MSU comparison. It consists of short scenes which quickly switch and a lot of special effects, such as red sparkles and grain. NIQE shows unnatural results on this sequence for all use cases: the rate-distortion curve is not monotonic because of an anomaly big values on dark frames. 


The videos described above contained completely black or dark frames. In these videos, NIQE had large values mostly on these frames, which was the main reason for the wrong overall quality score for the entire video. The following examples demonstrate another case in which NIQE was not applicable to video quality estimation.

\subsubsection{Noisy scenes/scenes with lots of details}

\begin{figure}[H]
    \begin{minipage}[b]{.98\linewidth}
    \centering{\includegraphics[width=\linewidth]{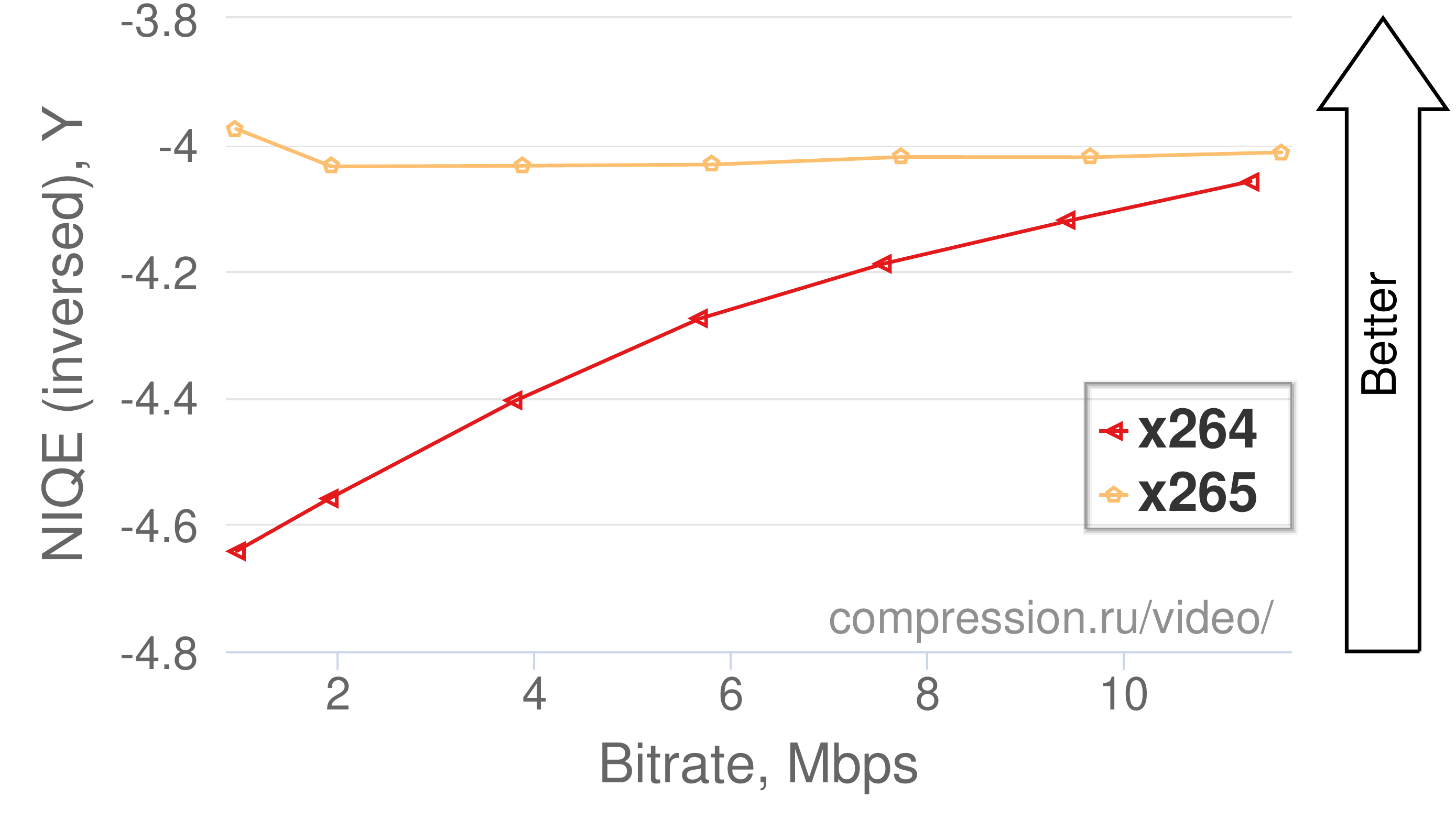}}
    \caption{Rate-distortion graph for \textit{Bay time lapse}.}
    \label{fig:bay_timelapse}
    \end{minipage}
\end{figure}

A number of cases where the metric took wrong values appear in videos with noise or a lot of small and textured details, like sand, water waves and grass. For x265-encoded \textit{Bay time-lapse} sequence, NIQE showed worse score on 2000 kbps than on 1000 kbps in universal use case (Fig.~\ref{fig:bay_timelapse}). This video contained a scene with water and grass, and the grass and waves on the water are smoother in a lower-bit rate video stream.

\begin{figure}[H]
    \begin{minipage}[b]{.98\linewidth}
    \centering{\includegraphics[width=\linewidth]{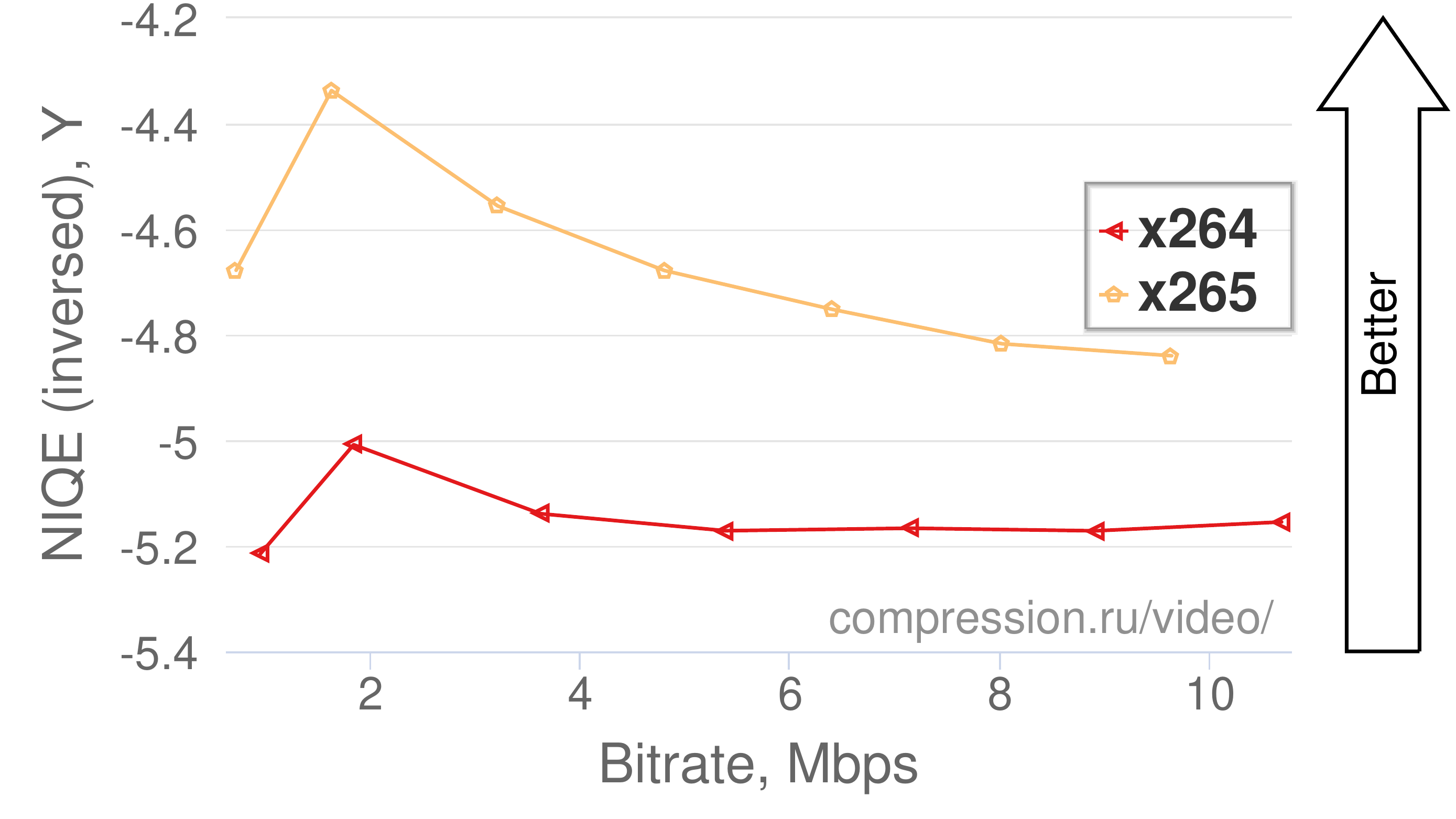}}
    \caption{Rate-distortion graph for \textit{Playground}.}
    \label{fig:playground}
    \end{minipage}
\end{figure}

In another example, NIQE showed worse score on 4000 kbps than on 2000 kbps in ripping use case on \textit{Playground} video sequence for both encoders. This video contains a lot of bright frames with highly structural and detailed grass and sand. Such texture is quite complicated for compression, and on low bit rates, there were visible compression artifacts, but NIQE had a worse score on high bit rates (Fig.~\ref{fig:playground}). This happened due to NIQE perception of finely textured grass as noise, while blurred compressed grass was expected to be visually better by NIQE. This is why the rate-distortion curve looks inverted on bit rates higher than 2000 kbps.

\subsection{Proposed processing technique}

During the analysis of per-frame NIQE results, it was revealed, that values greater than 40 don't usually appear in most of the video frames. Extreme values often occur in solid-colored or dark frames. We proposed and applied a special averaging technique to eliminate these cases. Our NIQE score for the video $V$ was computed in the following way:

\begin{equation}
\begin{aligned}
Score_V=\frac{\sum_{i} m_i*k_i}{\sum_{i} k_i}, i\in[0,N],
\\
k_i=\begin{cases}
    1, m_i\in[0,15),\\
    -0.04*m_i+1.6, m_i\in[15,40),\\
    0, m_i\in[40,+\infty), where
    \end{cases}
\end{aligned}
\end{equation}
\begin{expl}
    \item $m_i$ -- NIQE score for frame $i$,
    \item $k_i$ -- weighting coefficient for $m_i$ score,
    \item $N$ -- number of frames.
\end{expl}

The proposed averaging formula helped to improve NIQE scores for some of the video sequences. The following results demonstrate the corrected rate-distortion curves, which can be compared to the original results presented above.

\begin{figure}[H]
    \begin{subfigure}{.23\textwidth}
    \includegraphics[width=\linewidth]{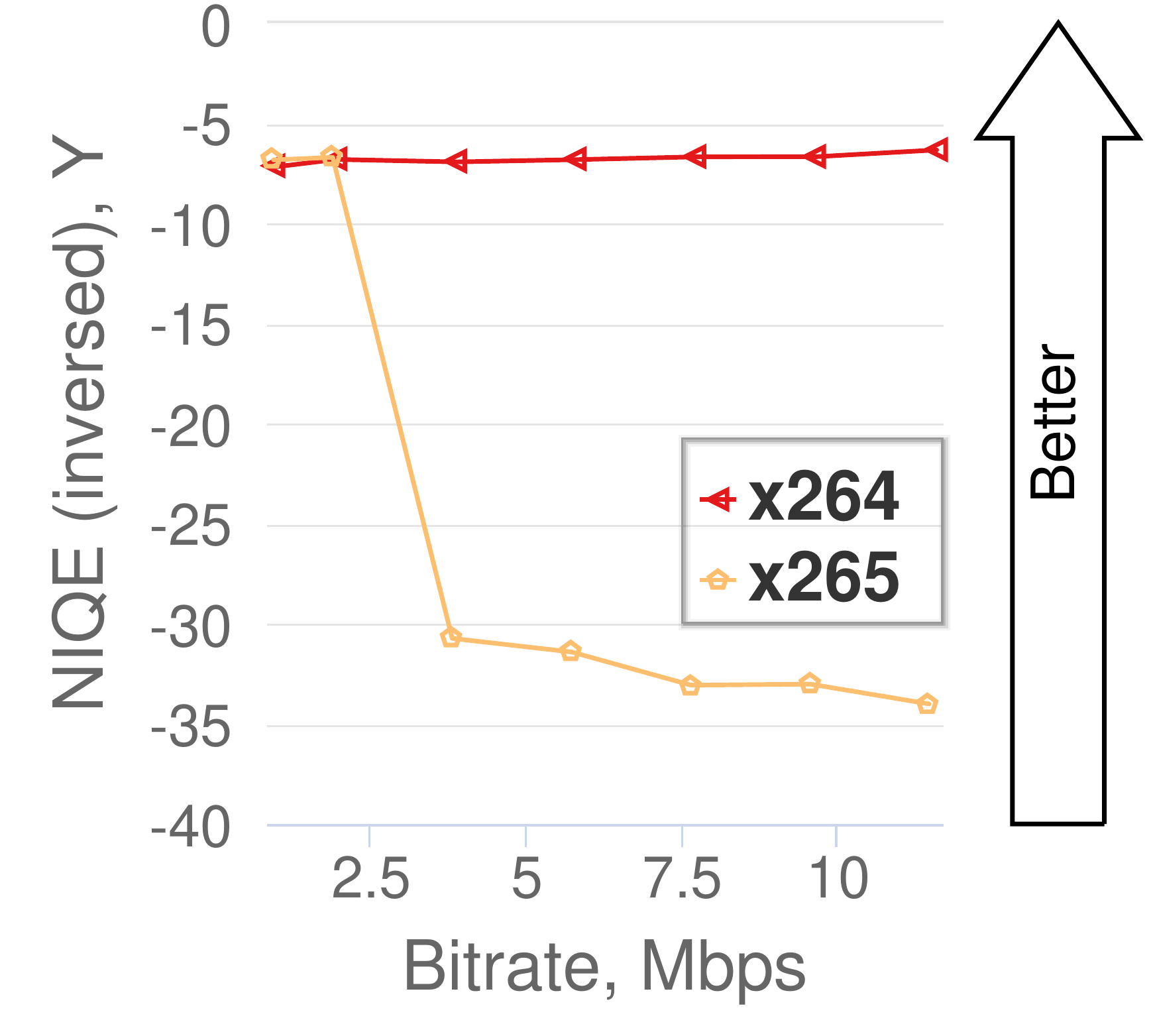}
    \caption{Original rate-distortion graph.} \label{fig:forest_dog_not_fixed}
    \end{subfigure}
    \hspace*{\fill}
    \begin{subfigure}{0.23\textwidth}
    \includegraphics[width=\linewidth]{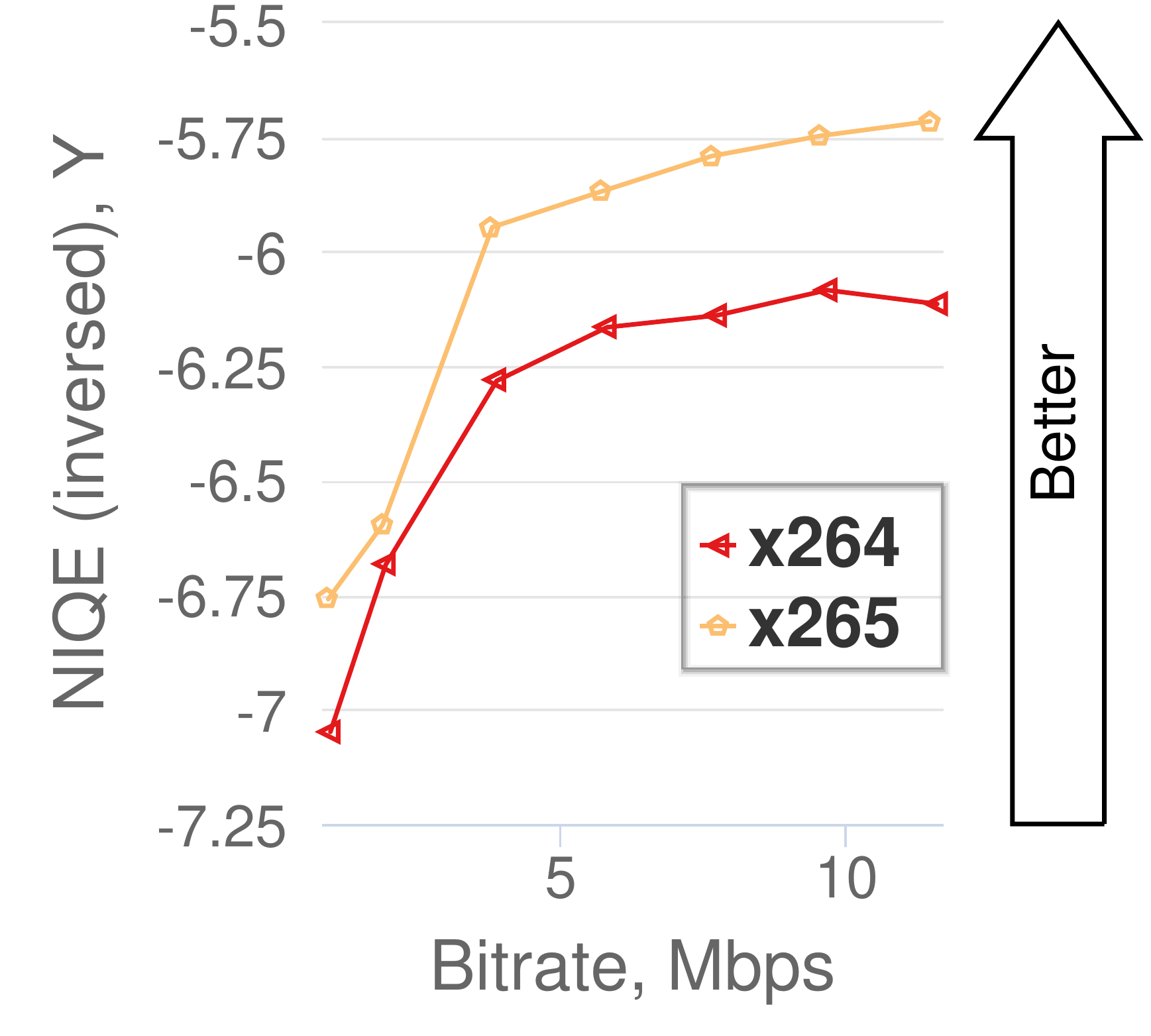}
    \caption{Rate-distortion graph after smart averaging.} \label{fig:forest_dog_fixed}
    \end{subfigure}
    \caption{\textit{Forest dog} video sequence.} \label{fig:forest_dog_comparison}
\end{figure}

\begin{figure}[H]
    \begin{minipage}[b]{.98\linewidth}
    \centering{\includegraphics[width=\linewidth]{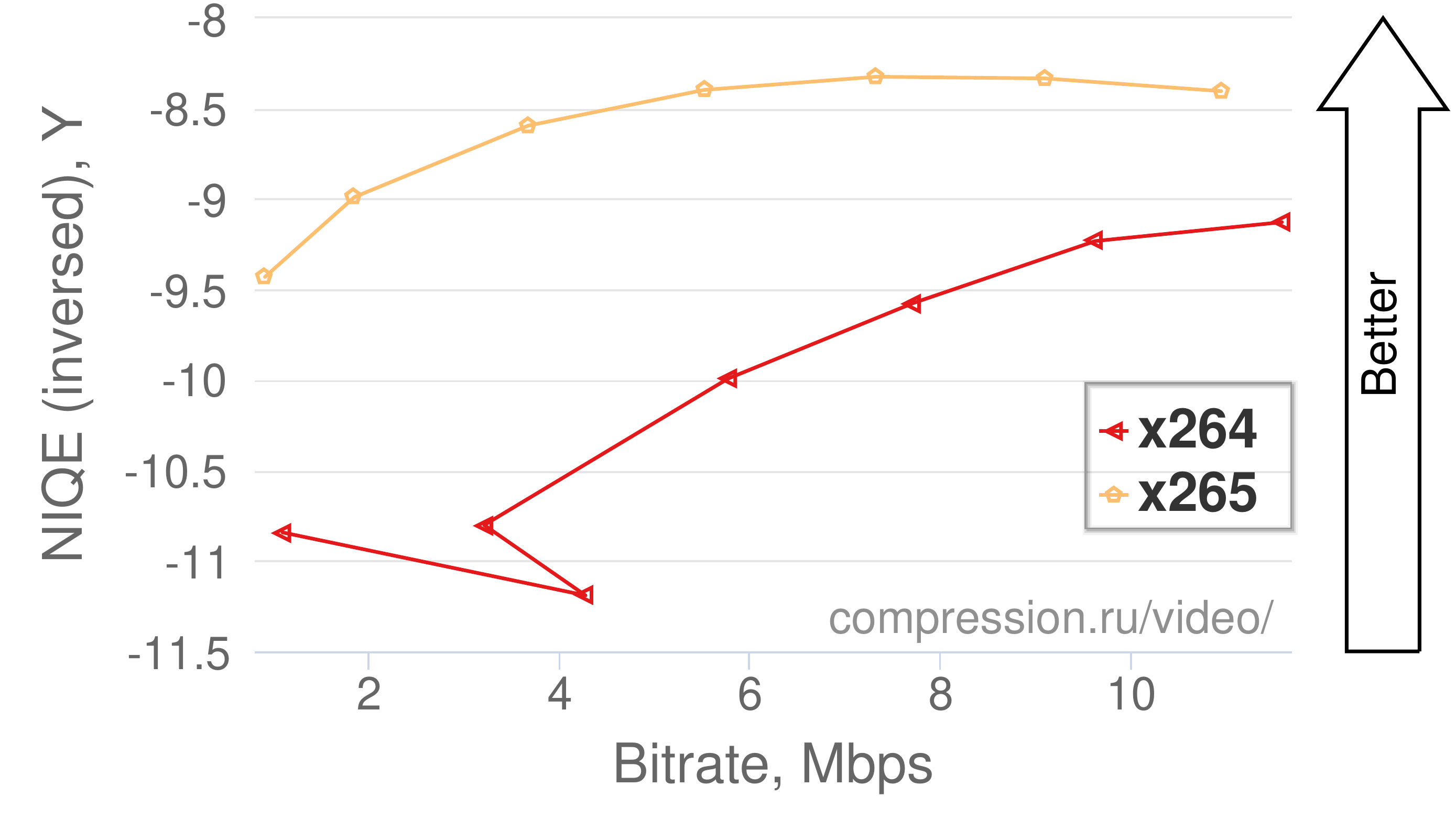}}
    \caption{Rate-distortion graph for \textit{Music clip} after smart averaging.}
    \label{fig:music_clip_fixed}
    \end{minipage}
\end{figure}

With the proposed averaging technique rate-distortion curve for \textit{Forest dog} doesn't contain outlying points (Fig.~\ref{fig:forest_dog_fixed}). 
Another example, where the results were corrected by the proposed averaging for both encoders, is \textit{Music clip} video sequence (Fig.~\ref{fig:music_clip_fixed}). The non-monotonic curve of x264 encoding was caused by high spatial complexity of this video.


\section{Correlation with subjective scores}

\begin{figure}[H]
    \begin{minipage}[b]{.98\linewidth}
    \centering{\includegraphics[width=\linewidth]{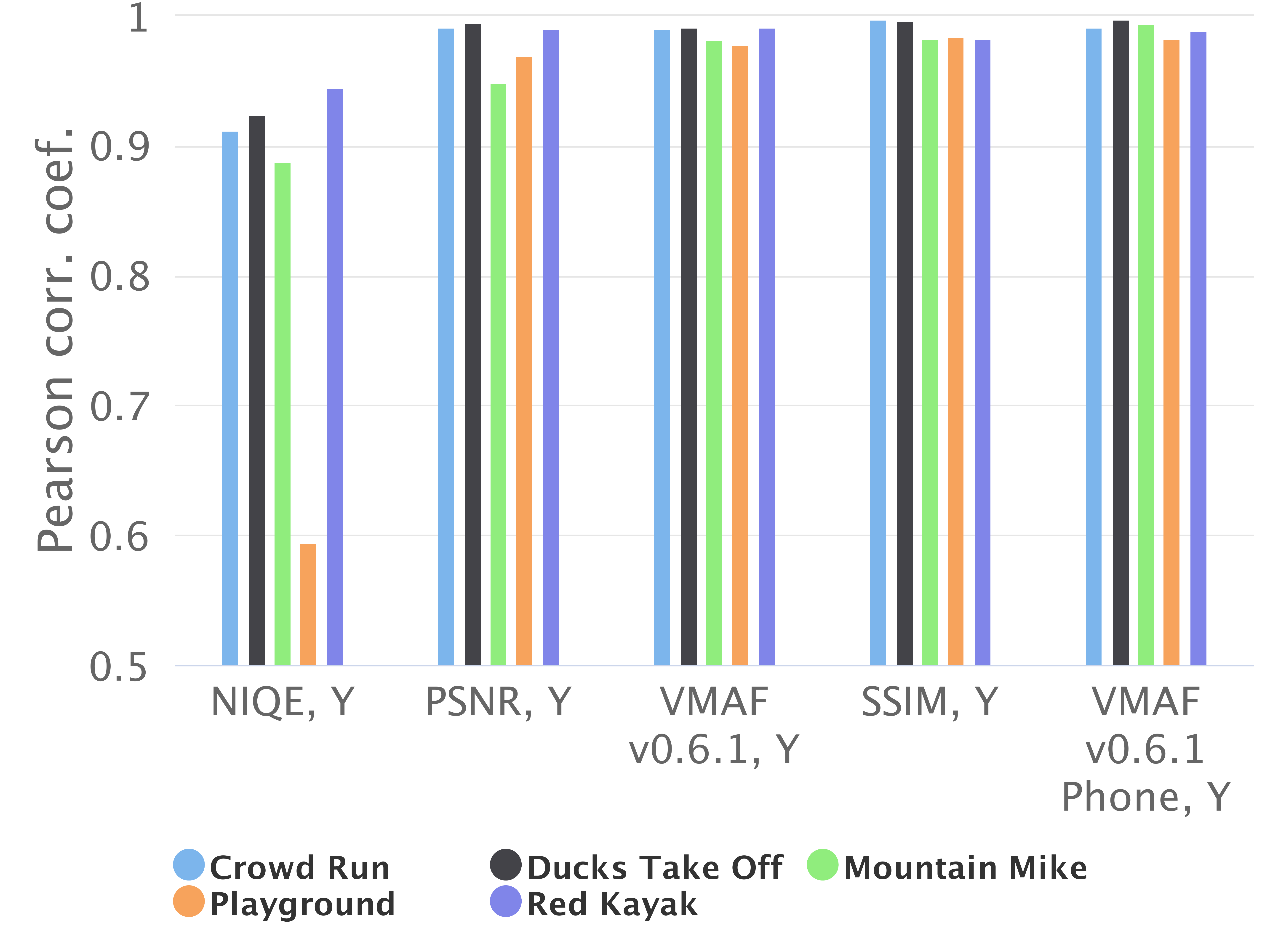}}
    \caption{Correlation between objective quality metrics and subjective scores.}
    \label{fig:metrics_corr}
    \end{minipage}
\end{figure}

The obtained NIQE quality scores were compared to the subjective scores on part of test videos. A pairwise subjective comparison was conducted as one of the parts of 2018 MSU Video-Codec Comparison, where a total of 22542 valid answers were received from 473 subjects. The detailed description and methodology can be found in the report \cite{msu_subjective_2018}. Five videos were used in this comparison, and none of them contained animated scenes or black frames for which NIQE could show inaccurate results. In addition, several full-reference quality metrics were measured (SSIM, PSNR, VMAF and their variations). The Pearson correlation coefficient was calculated for the results on each video separately (Fig.~\ref{fig:metrics_corr}). The averaged correlation scores across all videos reveal that NIQE has the lowest correlation with subjective scores (0.85) while VMAF v.0.6.1 for phones has the highest correlation (0.99). It should also be noted that at the moment NIQE has even lower correlation to subjective quality than PSNR (0.98), which is long considered to have low similarity to subjective quality for compression algorithms comparison.

\begin{figure}[H]
    \begin{minipage}[b]{.98\linewidth}
    \centering{\includegraphics[width=\linewidth]{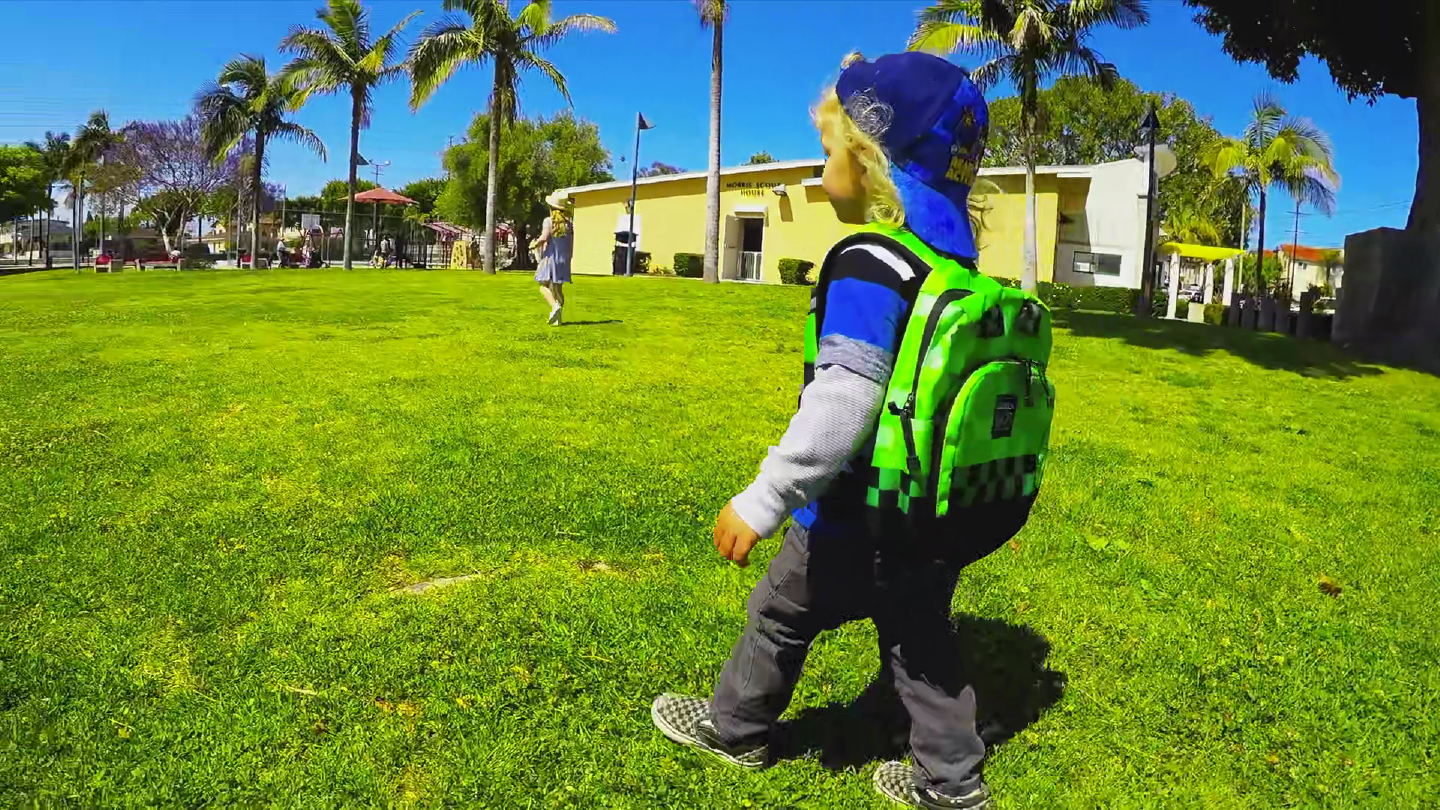}}
    \end{minipage}
    \\
    \begin{minipage}[b]{.98\linewidth}
        \begin{minipage}[b]{.495\linewidth}
        \centering{\includegraphics[width=\linewidth]{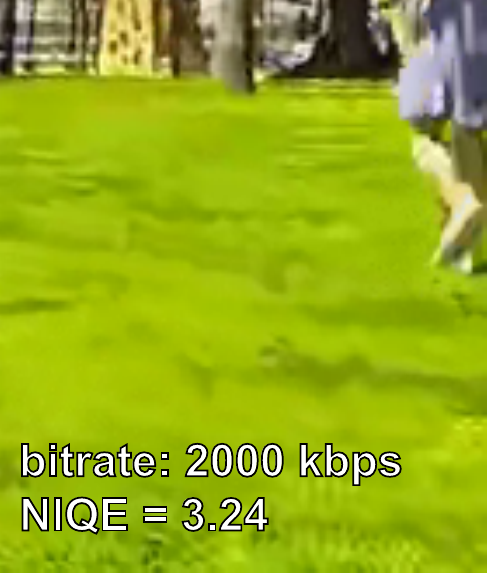}}
        \end{minipage}
        \hfill
        \begin{minipage}[b]{.495\linewidth}
        \centering{\includegraphics[width=\linewidth]{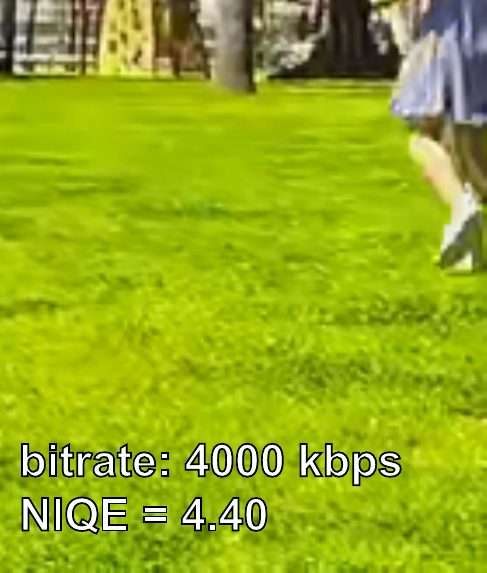}}
        \end{minipage}
    \end{minipage}
    \caption{Frame 58 from \textit{Playground} video sequence, codec: x265, ripping use case. According to NIQE, left image is visually better.}
    \label{fig:playground_vis}
\end{figure}

The lowest correlation of NIQE with subjective scores was obtained for \textit{Playground} video sequence. As it was described above for this video sequence, NIQE showed worse scores for detailed textures (grass and sand) in this video sequence, which is illustrated in Fig.~\ref{fig:playground_vis}.

\section{Conclusion}

During the experiments, NIQE showed good results for most of the videos. But still, there are many cases for which the metric is not applicable. This is why NIQE is not universal and can not be used in video-codec comparisons at the moment. The results of this comparison show NIQE deficiencies that need to be corrected, such as an application to animated cartoons, videos with completely black and solid-colored frames, noise and highly detailed/textured frames. For example, the abundance of fine details (grass, sand, grain effects) increases the values of NIQE despite the high bit rate of the encoded video, which leads to incorrect results. At the same time, in the original paper, NIQE was said to be not applicable to computer graphics, but in our investigation, it was found that the metric works for some types of animation (particularly for a screen capture of video gaming). 

\section{Acknowledgments}

Special thanks to Georgiy Osipov who helped to analyze all detected issues and improved NIQE implementation in MSU VQMT. This work was partially supported by the Russian Foundation for Basic Research under Grant 19-01-00785a.

\aboutAuthors

\end{multicols*}

\end{document}